# Multidimensional Design of Metal-Nitrogen-Carbon Electrocatalysts for Direct Propylene Epoxidation

Songbo Ye [a, b, †], Qingyuan Han [c, †], Jingwen Chi [c], Yuan Huang [c], Heng Liu [a], Di Zhang [a], Hitoshi Shiku [b], Li Wei [c, *], and Hao Li [a, *]

[a.] Advanced Institute for Materials Research (WPI-AIMR), Tohoku University, Sendai 980-8577, Japan

[b.] Graduate School of Engineering, Tohoku University, 6-6-11 Aramaki-aza Aoba, Aoba-ku, Sendai 980-8579, Japan

[c.] School of Chemical and Biomolecular Engineering, The University of Sydney, Darlington, New South Wales, 2006, Australia

[†.] These authors contributed equally: Songbo Ye, Qingyuan Han.

**Email**

l.wei@sydney.edu.au (L.W.)

li.hao.b8@tohoku.ac.jp (H.L.)

## Abstract

Propylene oxide (PO) is a major industrial chemical whose production currently relies on hazardous chlorine- or peroxide-based oxidants. Direct electrochemical epoxidation using water as the oxygen source offers a sustainable alternative, but controlling oxygen-atom transfer against the competing oxygen evolution reaction remains a fundamental challenge. Here, we show that propylene epoxidation selectivity cannot be described by oxygen binding energy alone, but is jointly governed by oxygen adsorption, the potential of zero charge (PZC), and applied potential. By combining theoretical calculations with pH-field coupled microkinetic modeling across 41 metal-nitrogen-carbon single-atom catalysts, we first identified an optimal oxygen-binding window and Co as the most favorable metal center. We then found that peripheral substituents can tune the PZC while largely preserving the optimal oxygen adsorption energetics, thereby providing an independent design dimension to further optimize the already favorable Co active site. This sequential, multidimensional design strategy identified CoPc-$NH_2$/CNT as the optimal catalyst, delivering a record PO Faradaic efficiency of 70–80% for direct propylene epoxidation in aqueous electrolyte under ambient conditions. These results establish interfacial electrostatics as an independently tunable design dimension for controlling selective oxygen-atom transfer in electrocatalysis.

## Introduction

Propylene oxide (PO) is a key industrial chemical primarily utilized in the manufacture of polyether polyols for polyurethane plastics,[1-3] with global production nearing 12 million tons in 2025, expected to reach 18 million tons by 2035.[4] To date, industrial PO production primarily relies on two thermo-catalysis approaches, i.e., the chlorohydrin and hydroperoxide processes, where molecular chlorine ($Cl_2$) and hydrogen peroxide ($H_2O_2$) are used as oxidants, respectively (**Fig. 1a**).[5] These processes are economically costly, produce environmentally hazardous chemical waste, and have a high risk of peroxide explosion.[6] More abundant oxidants such as molecular oxygen ($O_2$) can also be used to produce PO, but the processes are much less efficient in addition to the formation of significant amounts of by-products.[7]

Electrochemical propylene epoxidation driven by renewable electricity is a green and promising route, generating PO at the anode and green hydrogen at the cathode (**Fig. 1b**). Chloride can be oxidized at the anode to generate $Cl_2$, which acts as a redox mediator for epoxidation and enables high current density electrochemical chlorohydrin production.[8] However, like $Cl_2$-based thermal catalytic routes, chlorine-containing by-products raise significant environmental concerns.[9] Direct oxidation using water is an alternative method, instead of using external or generated $O_2$, it can utilize the atomic oxygen intermediate (O*, where * represents the adsorbed state on the anode surface) generated during the oxygen evolution reaction (OER) in an electrolyzer at room temperature for epoxidation. Previous studies suggested that $\mathrm{O}* + \mathrm{H_2O(l)} \leftrightarrow \mathrm{HOO}* + \mathrm{H^+} + \mathrm{e^-}$ can be a rate-determining step on many weak-binding catalysts.[10] Therefore, the stabilization of the O* intermediate on the catalyst in a certain potential range can afford an exciting avenue for propylene epoxidation.[11, 12] This electrocatalytic refinery ("*e*-refinery") strategy can potentially revolutionize the conventional resource-consuming and pollution-carrying epoxide industry.

**However, a main problem is that electrolytic propylene epoxidation has received limited attention in research over the past decades.** While a few pioneering studies have reported the use of Au, Ag, $PdO_x$, and $PtO_x$ as the anode catalysts for PO formation, follow-up studies were sparse after the 2000s, and the reported Faradaic efficiencies (FEs) for propylene were generally low (below 40%).[13] A major breakthrough was achieved recently with Pd-Pt oxide catalysts, which delivered substantially enhanced FEs of 60–70% for propylene epoxidation in water-acetonitrile electrolytes.[14]

Subsequent efforts have extended the investigation to a broader range of metal oxides, including AgFeOOH,[15] Ag-V-O,[16] amorphous perovskite-type oxides.[17] However, the FE for PO once again reaches a bottleneck (below 40%). Metal-based catalysts and their oxides seem to face intrinsic limitations. Multiple surface-active sites and potential-driven reconstruction under working conditions, including phase transitions and coverage changes, govern the selectivity of propylene epoxidation on Pd and Pt catalysts, potentially lowering PO selectivity (**Fig. 1c**).[18, 19]

To mitigate the complexity and selectivity issues associated with dynamically evolving surface states, herein, we turn to metal-nitrogen-carbon (M-N-C) single-atom catalysts (SACs).[20] The stable surface states of these catalysts are advantageous for improving PO selectivity.[21] Additionally, tailoring the central metal atoms (M = Co, Fe, Ni, *etc.*) and surrounding substituents (-H, -$NH_2$, tBu, *etc.*) enables the design of promising catalysts (**Fig. 1d**).[22, 23] However, designing PO-selective catalysts for propylene epoxidation remains challenging due to the tremendous cost of a conventional trial-and-error process in experiments. To address this challenge, theoretical calculations (e.g., density functional theory, DFT) and catalysis theory based on microkinetic modeling emerged as powerful tools that can provide guidelines for catalyst design.[24] Our previous studies demonstrated that pH-field coupled microkinetic modeling by considering interfacial factors including potential of zero charges (PZCs)[25] have provided design guidelines and mechanistic insights into various electrocatalytic reactions (e.g., oxygen and $CO_2$ electrocatalysis).[26-28] Such a strategy offers the potential to establish a precise theoretical screening model for identifying catalysts with high PO selectivity.

Herein, we aim to derive an electric field-coupled microkinetic model and identify qualitative relationship between PO selectivity in experiments using key descriptors (e.g., oxygen binding energy, applied potentials, and electric field effects). To achieve this, we performed DFT calculations to evaluate the binding energies on 41 M-N-C SACs with diverse central metals (e.g., Ti, V, Cr, Mn, Fe, Co, Ni, and Cu) and coordination environments (e.g., pyrrolic, pyridinic, MPc-R/CNT (R = $NH_2$, tBu, H, COOH, and F)). Then, we experimentally evaluated a series of MPc-H/CNT catalysts with different central metal atoms (M = Co, Cu, Fe, Ni, and Mn), which exhibit significantly varied O* binding energies, and measured their corresponding FEs for propylene epoxidation. After identifying Co as the optimal metal center, we then systematically modified the substituents of CoPc-R/CNT (R = $NH_2$, tBu, H, COOH, and F) to tune the electric field/PZC effects and examined the resulting changes in FE. The differences in PZC were identified using a solvation model.[25, 29] Within this framework, we construct a multidimensional model describing the competition between propylene epoxidation and OER, enabling qualitative description of PO selectivity across a multidimensional descriptor space. More importantly, as shown in **Fig. 1e**, we successfully designed and synthesized a non-precious-metal M-N-C catalyst, CoPc-$NH_2$/CNT, that achieves the highest FE (70–80%) and turnover frequency (TOF) reported to date in aqueous electrolytes.

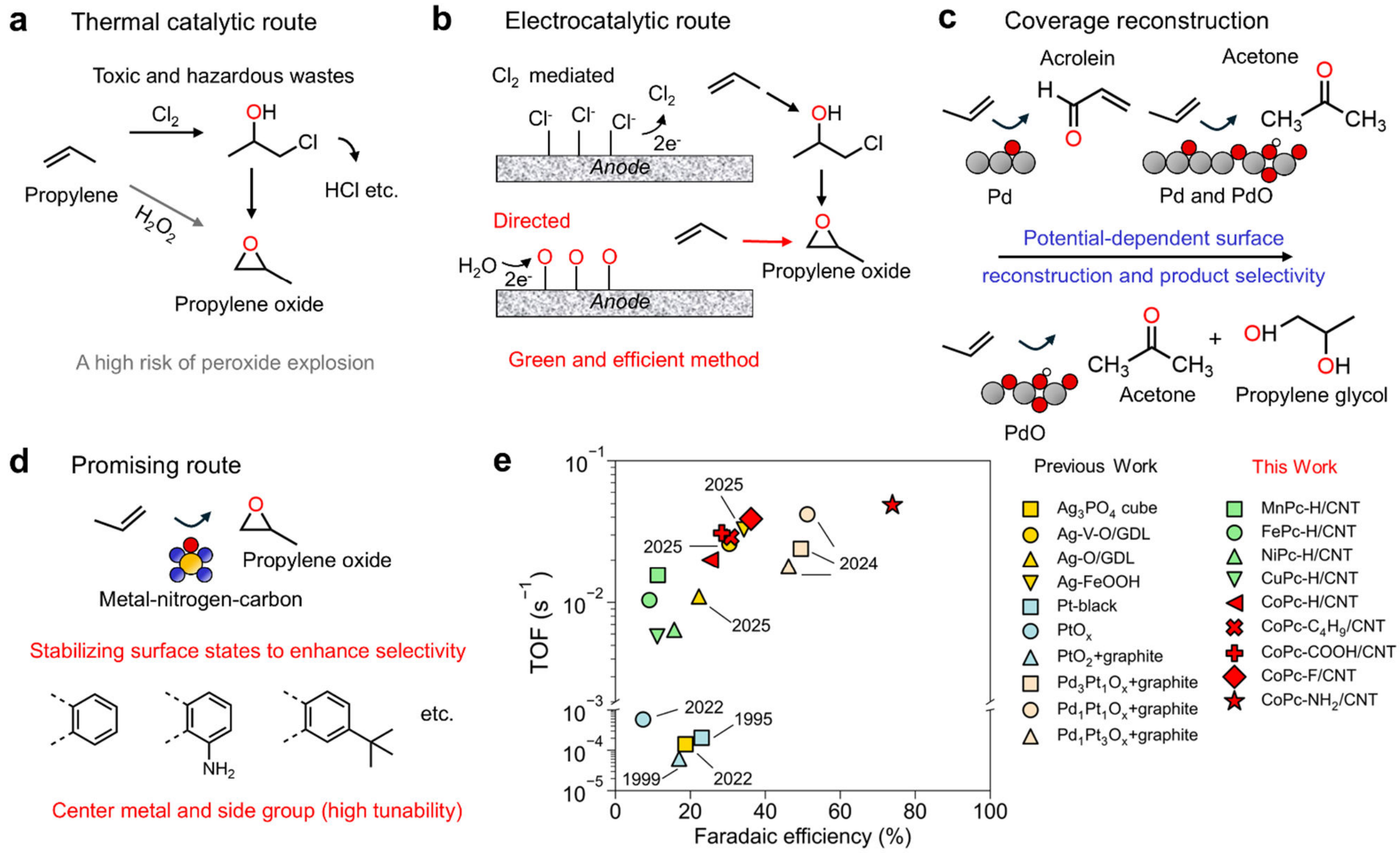

**Fig. 1. PO synthesis routes and performance of direct electrocatalytic propylene epoxidation. a,** Thermal catalytic propylene epoxidation and its limitations and **b,** electrocatalytic propylene epoxidation route. **c,** Potential-induced surface state change and the subsequent impact on product selectivity for metals and metal oxides as exemplified by Pd(111). **d**, Well-defined active-site states and tunable design space in M-N-C catalysts. **e,** Summary of the performance of reported direct electrochemical propylene epoxidation catalysts and this work. All performances were evaluated and compared in aqueous electrolytes.

# Results

## Oxygen binding energy as a key descriptor

DFT calculations for propylene epoxidation and OER were performed based on the proposed reaction mechanism (**Fig. 2a**). First, electrochemical water oxidation generates O* (*via* the first two proton-electron transfer steps, I and II). Subsequently, propylene and $H_2O$ compete for the O*, generating PO (II→V) and $O_2$ (II→III→IV), respectively. Previous research has suggested that the activity and selectivity predictions are likely governed by oxygen (O*) binding energy.[11] Therefore, we first focused on developing the microkinetic models of propylene epoxidation and OER using O* binding energy as the key descriptor. We performed DFT to evaluate the O* binding energies over 41 M-N-C SACs with diverse central metals (i.e., Ti, V, Cr, Mn, Fe, Co, Ni, and Cu) and coordination environments (i.e., pyrrolic, pyridinic, and MPc-R/CNT). The scaling relationships between the binding energies of HO* and O* (**Fig. 2b**), as well as between HOO* and HO* (**Fig. 2c**), were summarized. To further demonstrate the universality of these scaling relations, we also incorporated binding energy data for >200 metal oxide structures from our Digital Catalysis Platform (*DigCat*) database.[30] Estimates of the energy barriers for the elementary steps are based on the scaling relations reported by Nørskov and co-workers.[10] The scaling relationship between the energy barrier for propylene ($C_3H_6$) reacting with adsorbed O* and the O* binding energy on pyridinic and pyrrolic coordination structures (**Fig. 2d**), were calculated.

As a first, intuitive analysis, the complete microkinetic volcanoes at 2.0 $V_{RHE}$ (RHE: reversible hydrogen electrode) for both propylene epoxidation (I→II→V→I) and OER (I→II→III→IV→I) based on respective reaction mechanisms (**Fig. 2a**) were derived as shown in **Fig. 2e**. For both propylene epoxidation and OER on M-N-C SACs, there is generally one import step that determine the rate, the third step (V, III, respectively). This leads to a volcano-shaped dependence in which M-N-C SACs with O* binding energies on the left-leg (strong O adsorption) are limited by a too-strong O bonding that leaves kinetic limitations in O* desorption to $C_3H_6O$ and O* protonation to form HOO* for catalysis. This includes Fe/MnPc-H/CNT. A weaker O* bonding on the right-leg suppresses O* coverage formation, limiting the rate (e.g, on Ni/CuPc-H/CNT). To illustrate this mechanistic transition, we also analyzed the rate constants, $k_1$ ($C_3H_6$→$C_3H_6O$) and $k_2$ (O*→HOO*), without considering O* coverage (dashed lines in **Fig. 2e**), clearly showing that O* coverage influences the rate-determining kinetics. Notably, for propylene epoxidation, the onset of O* coverage influence (the intersection of the solid and dashed lines in **Fig. 2e**) does not lead to a distinct volcano maximum due to the low rate constant $k_2$. This indicates that, for CoPc-H/CNT, the turnover frequency (TOF) of OER is limited by the O* coverage, whereas the TOF of propylene epoxidation is jointly limited by both the rate constant and the O* coverage. $k_2$ shows a stronger dependence on O* binding energy than $k_1$ (steeper slope), increasing rapidly as the O* bonding weakens. A maximum rate emerges at a second inflection point as the O* binding energy decreases further. This leads to a discrepancy in the volcano maxima, with propylene favoring weaker O* bonding than OER. This also identifies a range of O* binding energies (yellow region) that are favorable for PO production. Additionally, kinetic volcanoes were constructed for different voltages, at 2.2 $V_{RHE}$ (**Fig. 2f**) and 2.5 $V_{RHE}$ (**Fig. S1**) based on the computational hydrogen evolution (CHE) model.[31] As the $V_{RHE}$ increases, the favorable window of O* binding energy for PO formation gradually narrows. When the potential reaches 2.5 $V_{RHE}$, even with tuning of the O adsorption energy, it becomes nearly impossible to obtain catalysts with high PO selectivity. Since positive potentials promote O*, the onset of O* coverage decrease shifts progressively to higher O* binding energies, which in turn leads to a rightward shift of the intersection between the TOFs (solid line) and rate constants (dash line). Based on the adsorption energy screening, CoPc-H/CNT was selected for further study due to its favorable activity and selectivity in theory.

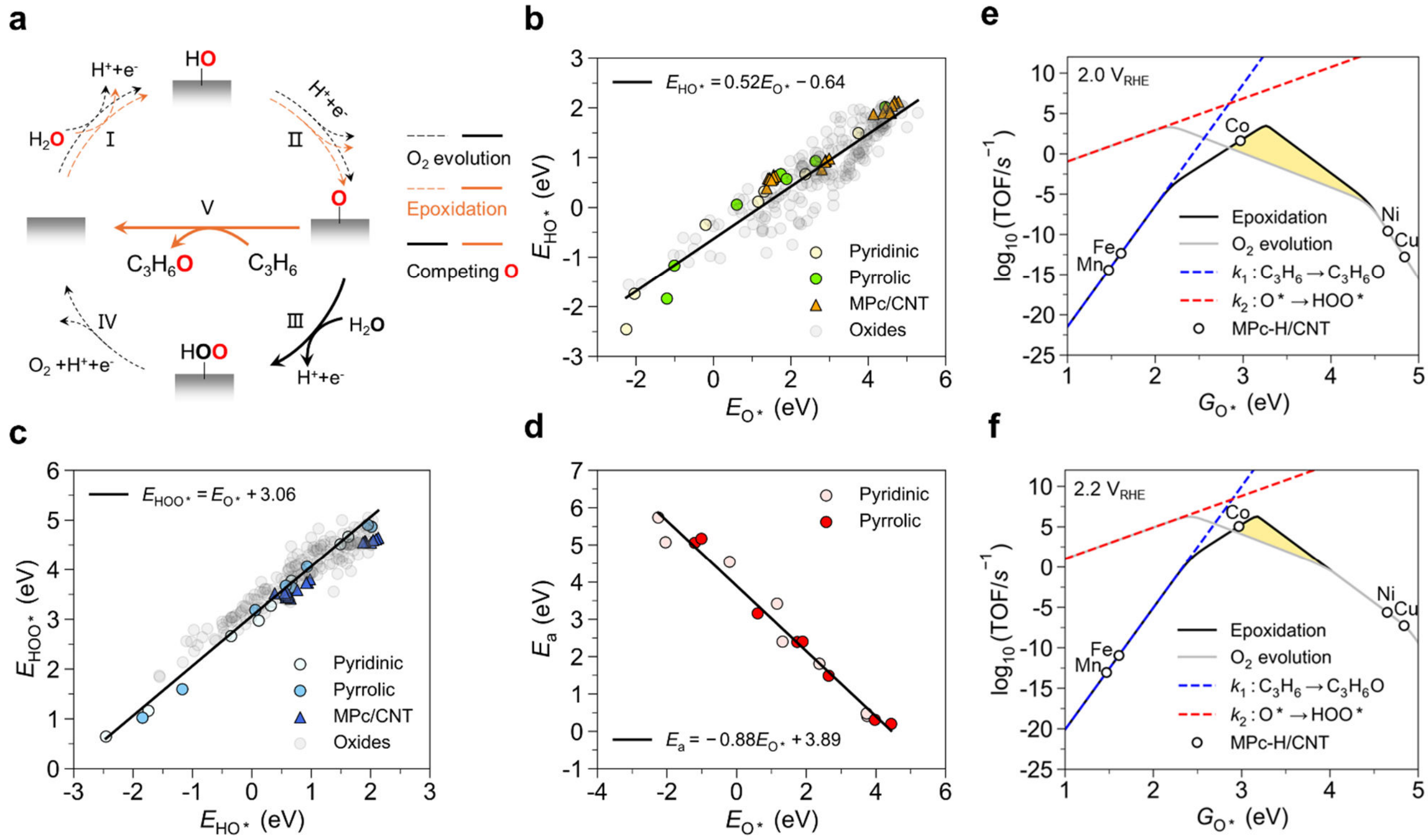

**Fig. 2. Linear scaling relations and microkinetic volcano models for the propylene epoxidation and OER process.** Proposed kinetic modeling mechanism and scaling relations in M-N-C SACs and oxides, including **a,** kinetic modeling scheme, **b,** $E_{HO*}$ *versus* $E_{O*}$ scaling relation, **c,** $E_{HOO*}$ *versus* $E_{HO*}$ scaling relation, and **d,** scaling relation between the activation barrier of propylene epoxidation

$E_a$ and $E_{O*}$, **e,** volcano at 2.0 $V_{RHE}$, **f,** volcano at 2.2 $V_{RHE}$. The solid line represents the theoretical TOF values ($s^{-1}$). The dashed lines denote the rate constants (k) of individual reaction steps derived under the steady-state approximation, neglecting surface O* coverage effects. The yellow region indicates that the epoxidation rate exceeds that of OER.

**Multidimensional design principles**

Building upon the O* binding energy as a primary descriptor, we aim to extend the analysis by incorporating the electric field/PZC into a multidimensional descriptor framework. Koper et al.[32, 33] and Li et al.[25] have identified the importance of the relationship between PZC and electrocatalytic reactions (Pt, SACs *etc*.). However, a limitation commonly remains: PZC and adsorbent binding energy ($H_2O$*, HO*, *etc*.) were theoretically considered to correlate with each other through an approximated linear relationship on transition metal surfaces.[34] This makes it difficult to determine whether the effect is dominated by adsorbent binding energy or by the PZC. This correlation also exists in SACs across different metal centers.[25] A promising strategy lies in tuning the PZC *via* substituent modification of MPc-H/CNT while keeping the adsorption energy nearly unchanged. Five MPc-R/CNT with different substituents (R = $NH_2$, tBu, H, COOH, and F) were selected, and their optimized structures are shown in **Fig. S3**. For different metal atoms, the effects of substituents on the PZC and work function (WF) are similar (**Fig. S4**). The PZC and WF exhibit a positive correlation (**Fig. S5**).[35] Using -H as reference, the effects of different substituents on the PZC and WF were evaluated by the average differences (**Fig. 3a**). Compared with -H, they tune both the PZC and WF across a broad range, from negative to positive shifts. The average effects of -$NH_2$ (PZC = −0.24 V, WF = −0.23 eV), -tBu (PZC = −0.03 V, WF = −0.15 eV), -COOH (PZC = 0.13 V, WF = 0.23 eV), and -F (PZC = 0.29 V, WF = 0.30 eV) on PZC and WF also exhibit a positive correlation. Moreover, the effects of substituents on binding energies of O*, HO*, and HOO* do not show an obvious influence (**Fig. S6**). All in all, by breaking the linear relationship between PZC and adsorbent binding energy, we can potentially distinguish between the effects of them.

To analyze the coupling effects of O* binding energy and PZC on PO selectivity and activity, we derived a microkinetic model for propylene epoxidation that also accounts for OER, in which all reaction pathways shown in **Fig. 2a** are included in this model. At 2.2 $V_{RHE}$ (**Fig. 3b**), CoPc-R/CNT can exceed OER in PO TOF by tuning the PZC *via* side groups. For Co/Fe/MnPc-R/CNT, a lower PZC enhances PO selectivity. In contrast, PZC shows a negligible influence on Cu/NiPc-R/CNT. For activity, a more positive PZC enhances the TOF of both PO formation (**Fig. 2c**) and OER (**Fig. S7**). The simulation results at different potentials show similar trends (**Fig. S8**). In addition, a more positive PZC shifts the volcano peak toward the right (i.e., weaker O* bonding). This is because, compared with HO*, O* exhibits a more sensitive response to the electric field (larger polarizability and dipole moment).[36] A positive PZC can enhance the O* binding, resulting in a larger O* coverage area (**Fig. S9**). To further understand the effects of PZC on PO selectivity and activity, we developed 1D microkinetic volcanoes of PO and OER under different potentials and PZC conditions (**Fig. S10**). The increase in O* coverage shows a clear correlation with OER activity, consistent with our description in **Figs. 2** and **3** of how potential affects OER activity as a function of O* coverage. Similarly, PZC influences OER activity by regulating O coverage. However, for PO formation, beyond the onset of O* coverage influence, PZC not only modulates the surface coverage but also alters the kinetic energy barriers of the PO pathway. For PO formation, between the first turning point and the volcano peak, the dual regulation of the PO rate by PZC, through both coverage and kinetic barriers, leads to a lower sensitivity compared with OER. Therefore, upon decreasing the PZC, the reduction in the PO rate is smaller than that of OER, which explains why a negative PZC is favorable for enhancing PO selectivity. To clearly understand the coupling effects between PZC and voltage, we developed a 2D microkinetic volcano (**Fig. S11**) with PZC and potential as dual descriptors and further investigated the variation of PO Faraday efficiency as a function of potentials (**Fig. S12**). Due to a change in the rate-determining step, on the left-leg of the PO volcano, simultaneously decreasing both the PZC and the potential is beneficial for enhancing PO selectivity. However, on the right-leg, selectivity is primarily governed by potential. On the too-strong-binding side of the volcano (e.g., $\Delta G_{O*} = 2.0$ eV, **Fig. S12a**), the PO Faraday efficiency is nearly zero. On the moderate binding energy region of the volcano, the PO Faraday efficiency can potentially reach 100% by adjusting the potential and PZC (e.g., $\Delta G_{O*} =$ 3.0 eV, 3.8 eV, **Figs. 12b, c**). On the too-weak-binding side of the volcano (e.g., $\Delta G_{O*} = 4.5$ eV, **Fig. S12d**), the PO Faraday efficiency decreases with potential and then stabilizes.

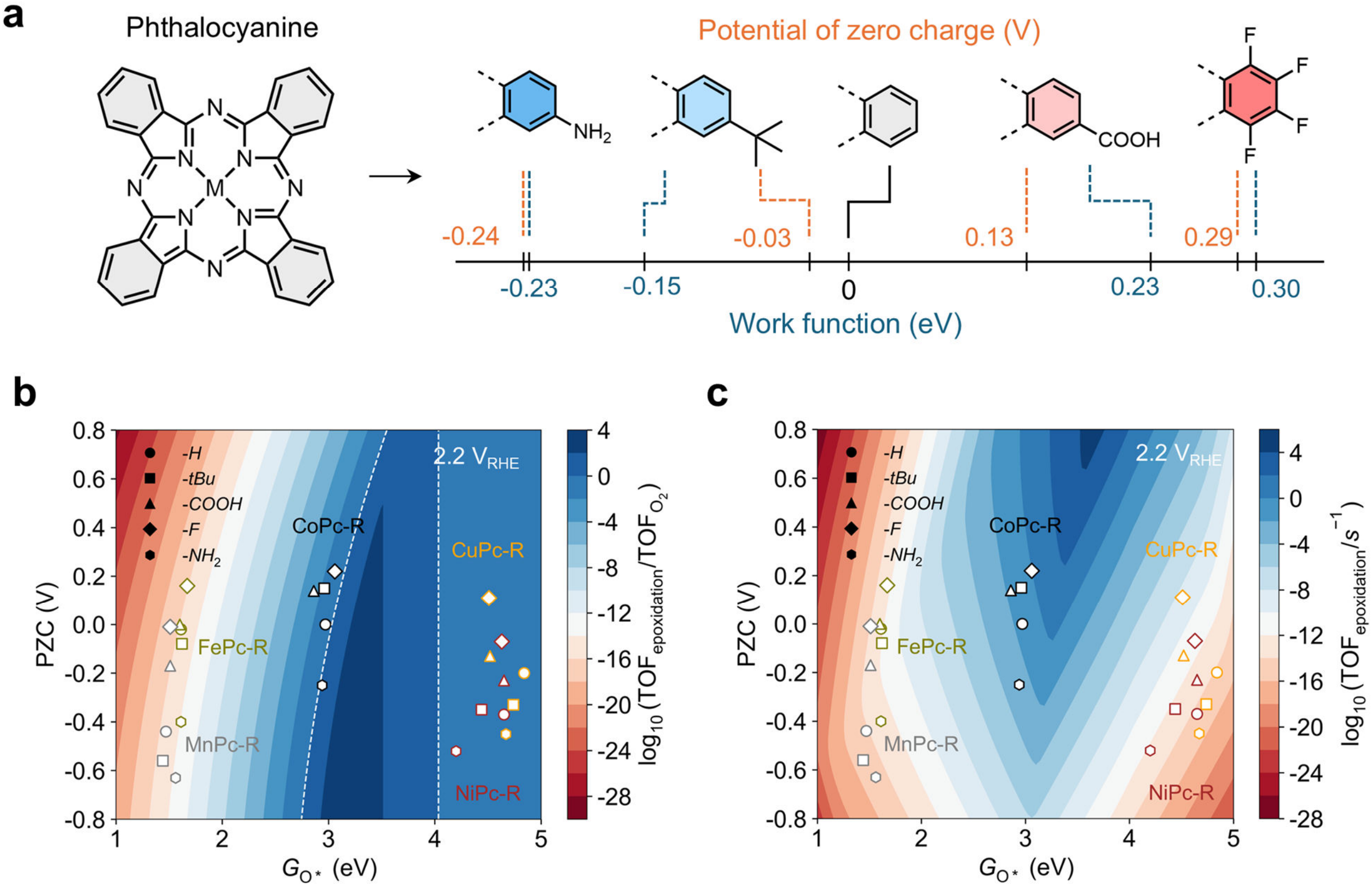

**Fig. 3. O* binding energy and PZC as descriptors for predicting the activity and selectivity of propylene epoxidation. a,** Substituted phenylene structures in MPc-R/CNT (M = Co, Cu, Fe, Mn, and Ni; R = $NH_2$, tBu, H, COOH, and F), and the average effects of different structures on PZC and WF relative to MPc-H/CNT as the reference, **b,** A two-dimensional selectivity volcano plot at 2.2 $V_{RHE}$ for propylene epoxidation, TOF, turnover frequency, **c**, A two-dimensional activity volcano plot at 2.2 $V_{RHE}$ for propylene epoxidation. The white dashed line indicates that the TOF for propylene epoxidation and OER are equal.

**Experimental validation and benchmarking**

To experimentally validate the multidimensional descriptor framework, a series of MPc-H/CNT (M=Mn, Fe, Co, Ni, and Cu) and CoPc-R/CNT catalysts (R = $NH_2$, tBu, H, COOH, and F) were synthesized, characterized, and tested for their performance in direct propylene epoxidation (see details in the **Method section** in the **Supporting Information**). Metal loadings determined by ICP-OES ranged from 0.2 to 0.9 wt% (**Table S1**) are found among different samples. High-angle annular dark-field scanning transmission electron microscopy (HAADF-STEM) imaging confirm that metal atoms are atomically dispersed on the CNT support with no observable metallic clusters or molecule aggregates and uniform elemental distribution under X-ray energy dispersive spectroscopy (EDS) elemental mapping (**Figs. 4a** and **S14**). The M-$N_4$ configuration in different catalysts was further supported by extended X-ray absorption fine structure (EXAFS) analysis (**Fig. S15**). Moreover, the CoPc-R/CNT catalysts exhibit varied edge-energy and white line peak features, indicating their varied Co electronic structure. This feature is also supported by the high-resolution Co 2p X-ray photoelectron spectra (XPS, **Fig. S16**).

Electrochemical propylene epoxidation was performed in a three-electrode H-cell with 0.5 M PBS electrolyte (pH 7.2) under continuous propylene flow by chronoamperometry testing. $^1$H NMR was used to assess liquid products using 2,2-dimethyl-2-silapentane-5-sulfonate (DSS) as an internal standard and confirmed PO formation (**Figs. S17-S18**). Acetic acid and hydroxyacetone were identified as the major by-products. The performance of a series of MPc-H/CNT catalysts (M = Mn, Fe, Co, Ni, and Cu) were evaluated firstly. Linear sweep voltammetry (LSV) curves obtained under Ar or propylene flow show varied current-potential responses (**Fig. S19**). The Faradaic efficiencies of different catalysts were quantified in the 1.8–2.58 $V_{RHE}$ window and compared in **Fig. S20**. CoPc-H/CNT exhibited the highest $FE_{PO}$, exceeding all other metal centers. Its maximum $FE_{PO}$ was approximately 1.6 times that of NiPc-H/CNT and 2.8 times that of FePc-H/CNT at their respective maximum $FE_{PO}$. FePc-H/CNT exhibited the lowest selectivity, whereas MnPc-H/CNT and CuPc-H/CNT showed intermediate performance. This trend is consistent with the theoretical microkinetic volcano model, in which the O* binding energies of these metal centers fall outside the favorable window for PO selectivity (**Fig. 2e**).

Based on this promising observation, the performance of a series CoPc-R/CNT catalysts was further assessed. Varied current responses were found from their LSV curves collected under Ar or propylene flow (**Fig. S21**). The $FE_{PO}$ determined across this potential window were further compared in **Fig. 4b**. Towards the lower end of window, all catalysts can produce measurable $FE_{PO}$ (**Fig. S22**). CoPc-$NH_2$/CNT achieved highest peak $FE_{PO}$ among all catalysts, approximately 3 times that of CoPc-H/CNT, which was also the highest reported to date for direct electrochemical propylene epoxidation in pure aqueous media under ambient conditions. No DMPO-trapped hydroxyl radical signal was detected in the anolyte of CoPc-$NH_2$/CNT under operating conditions (**Fig. S23**), indicating that freely diffusing hydroxyl radicals are not the active oxidant and pointing to a surface-confined oxygen-transfer pathway.

Moreover, the PO selectivity differences among substituents readily reflect a regime in which O* coverage remains the primary kinetic bottleneck. Specifically, a clear substituent-dependent divergence emerges toward the upper end of the potential window: CoPc-$NH_2$/CNT maintains significantly higher $FE_{PO}$ while the other catalysts exhibit a rapid decline, with CoPc-F/CNT showing the steepest decay. This potential-dependent behaviour is consistent with the microkinetic prediction that PZC becomes the dominant selectivity descriptor at elevated potentials, where a lower PZC preserves the kinetic advantage of the epoxidation pathway over OER. The experimentally observed selectivity ordering aligns qualitatively with the calculated PZC sequence ($NH_2$ < tBu < H < COOH < F). Further benchmarking against the microkinetic model reveals a linear correlation between the theoretically predicted and experimentally measured potentials required to reach a benchmark current density of 5 mA $cm^{-2}$ across the CoPc-R/CNT series (**Fig. 4c**), demonstrating the quantitative predictive power of the multidimensional descriptor framework for catalyst activity.

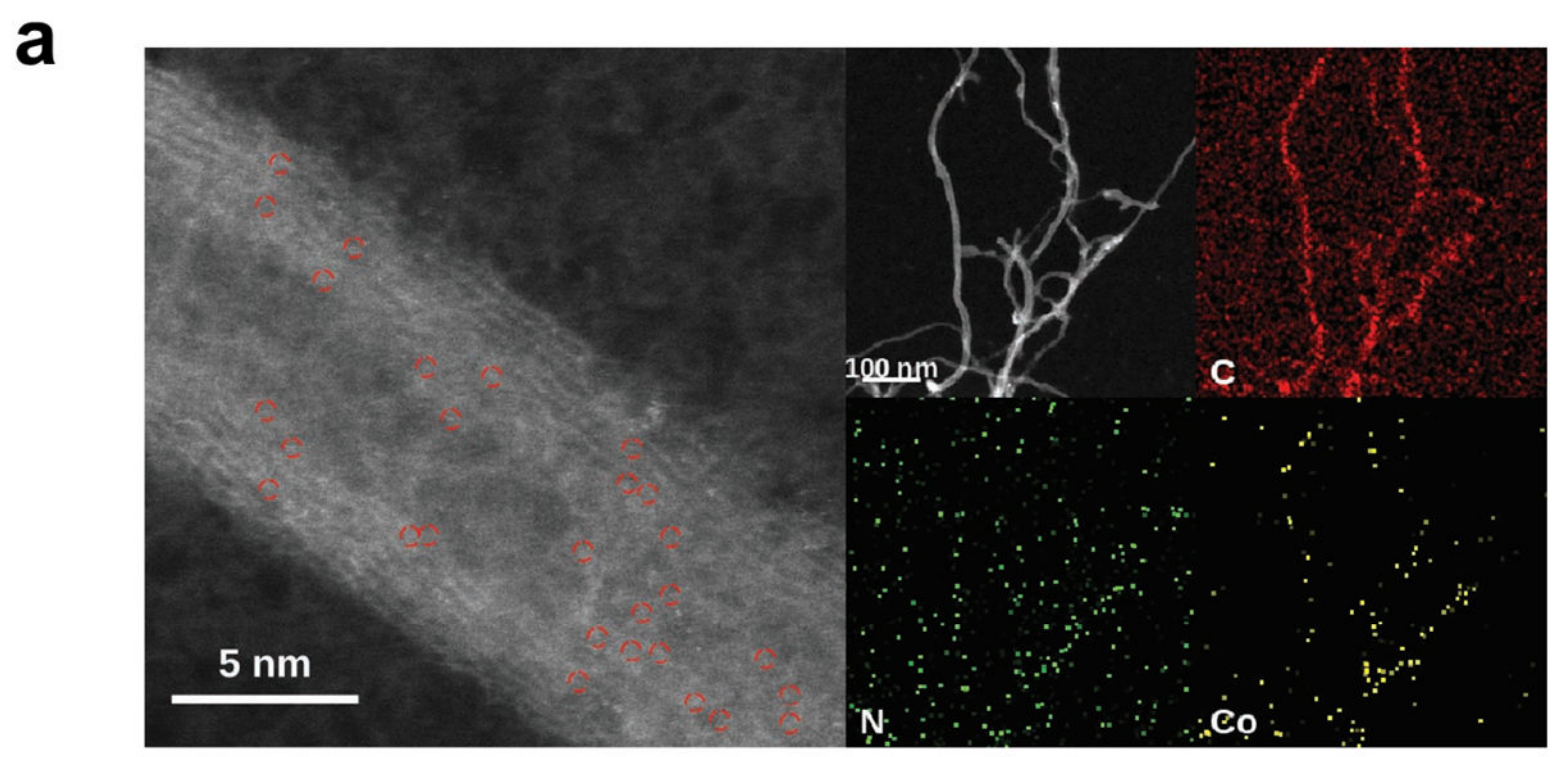

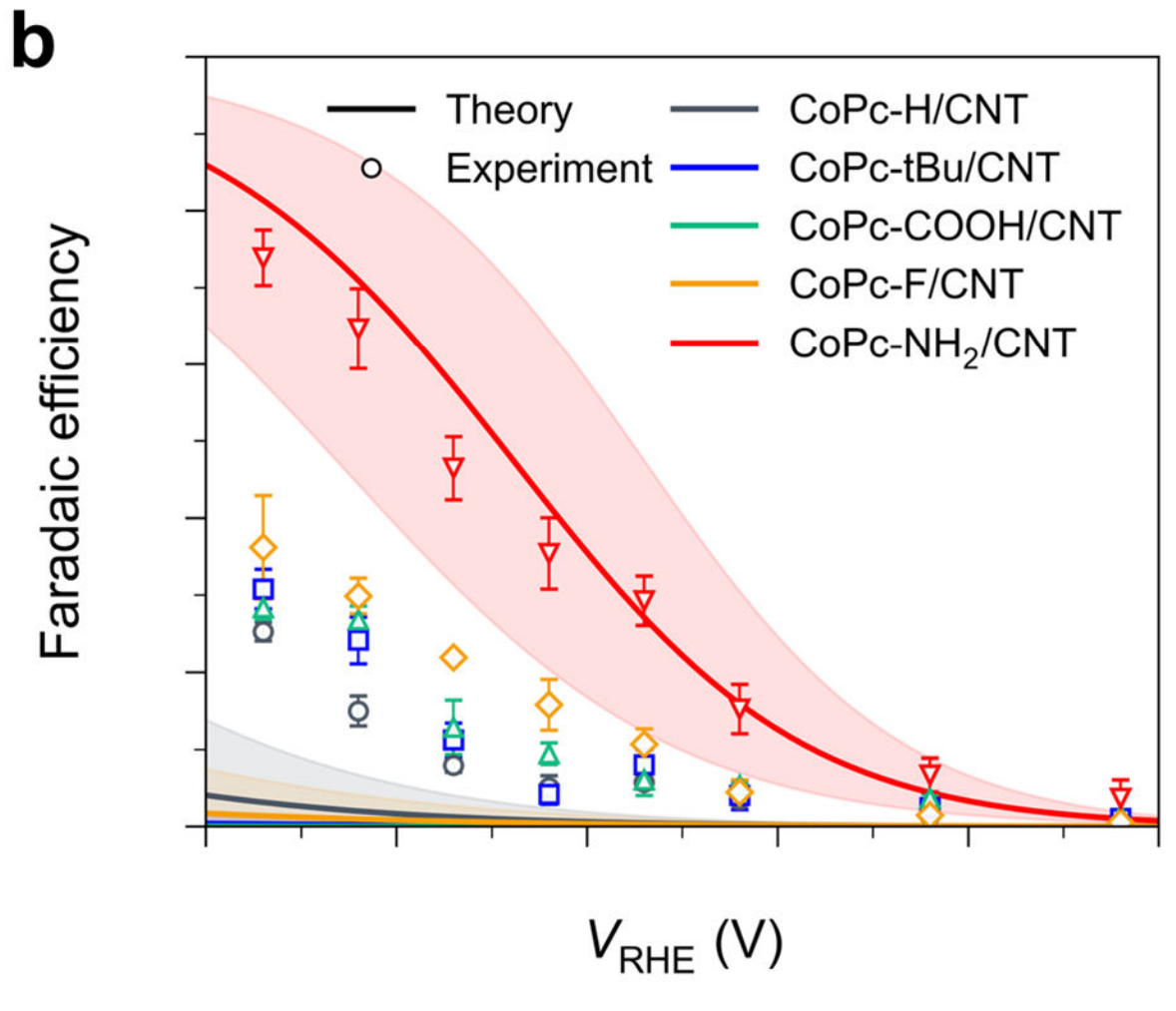

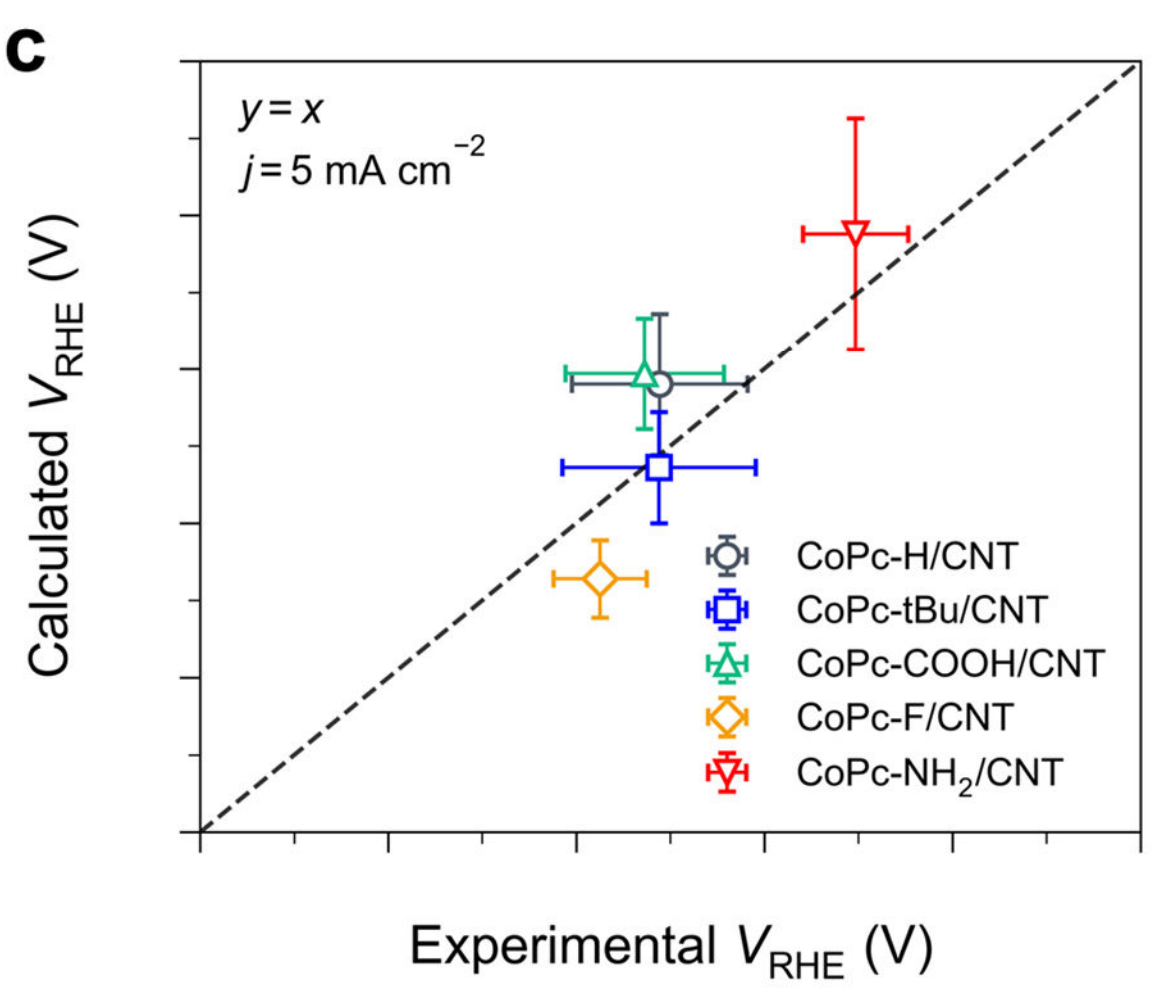

**Fig. 4. Experimental validation of the multidimensional design for PO production. a,** HAADF-STEM image and EDS elemental mapping of CoPc-$NH_2$/CNT catalyst. Atomically dispersed Co sites are highlighted in red circles. **b,** Comparison of experimental (open symbols) and theoretic (solid lines and shaded regions) $FE_{PO}$ as a function of applied potential ($V_{RHE}$) for CoPc-H/CNT, CoPc-tBu/CNT, CoPc-COOH/CNT, CoPc-F/CNT, and CoPc-$NH_2$/CNT catalysts. **c,** Comparison of calculated and experimental potential required to reach 5 mA $cm^{-2}$; the dashed line indicates the ideal 1:1 correlation**.**

## Conclusion

This work establishes a multidimensional framework for understanding and controlling direct electrochemical propylene epoxidation. We show that O* binding energy alone is insufficient to describe the competition between propylene epoxidation and the OER. Instead, selectivity is jointly governed by O* adsorption, PZC, and applied potential. Using O* binding energy as the first design dimension, we identified an optimal adsorption window and Co as the most favorable metal center among a broad range of M-N-C SACs. We then introduced the PZC as a second, independently tunable descriptor. By modifying the peripheral substituents of CoPc, the PZC could be systematically shifted while largely preserving the favorable O* adsorption energetics of the Co center. This enabled further optimization beyond metal-center selection and pushed CoPc-$NH_2$/CNT into the optimal selectivity regime. The resulting pH-field coupled microkinetic model reproduces the experimentally observed dependence of PO selectivity on metal center, substituent, and applied potential, while also capturing the activity trends across the CoPc-R/CNT series. Guided by this sequential design strategy, CoPc-$NH_2$/CNT achieves a record Faradaic efficiency of 70–80% and the highest reported turnover frequency for direct propylene epoxidation in aqueous electrolyte under ambient conditions. More broadly, this work demonstrates that interfacial electrostatics can serve as an independent design dimension beyond conventional adsorption-energy optimization, providing a general strategy for controlling selective oxygen-atom transfer in competition with water oxidation.

**Data Availability**

All the data are available from the authors upon request. The key experimental data, computational structures and energetics, and the microkinetic models are available in the Digital Catalysis Platform (*DigCat*, https://www.digcat.org/).

**Acknowledgements**

H.L. thanks the financial support from JSPS KAKENHI (No. JP23K13703). S.Y. acknowledges the JST-SPRING (JPMJSP2114). L.W. acknowledge the financial support from ARC under the Future Fellowship (FT210100218) and Linkage Projects (LP230200886). H.L. and L.W acknowledges the Center for Computational Materials Science, Institute for Materials Research, Tohoku University for the use of MASAMUNE-IMR (Project No. 202512-SCKXX-0218), the Institute for Solid State Physics (ISSP) at the University of Tokyo for the use of their supercomputers, Sydney Info Hub at the University of Sydney, and the National Computational Infrastructure (NCMAS-2024-59).

**Author contributions**

S.Y., H.L., L.W., and H.S. designed the study and wrote the manuscript. S.Y. conducted DFT calculations, data analysis, and microkinetic modelling. Q.H. performed the synthesis and electrochemical epoxidation experiments of the CoPc-R/CNT catalysts and drafted the experimental sections of the main text and SI. J.C. performed the electrochemical experiments and product characterization for the MPc-H/CNT metal-screening series. Y.H. performed materials characterization. S.Y., H.L., and D.Z. analyzed the theoretical data. All authors reviewed and edited the manuscript.

**Competing interests**

The authors declare no competing interests.

**Additional information**

The data that support the findings of this study are included in the published article and its **Supporting Information**. The following content is available free of charge. Details on the computational methods and of all simulated models; models of MPc-R/CNT, and details of binding free energy calculations; microkinetic modeling; supporting figures and tables.

## Author information

**Corresponding Author**

**Hitoshi Shiku -** Graduate School of Engineering, Tohoku University, 6-6-11 Aramaki-aza Aoba, Aoba-ku, Sendai 980-8579, Japan. Email: hitoshi.shiku.c3@tohoku.ac.jp

**Li Wei -** School of Chemical and Biomolecular Engineering, The University of Sydney, Darlington, New South Wales, 2006, Australia. Email: l.wei@sydney.edu.au

**Hao Li -** Advanced Institute for Materials Research (WPI-AIMR), Tohoku University, Sendai 980-8577, Japan. Email: li.hao.b8@tohoku.ac.jp

**Authors**

**Songbo Ye -** Advanced Institute for Materials Research (WPI-AIMR), Tohoku University, Sendai 980-8577, Japan; Graduate School of Engineering, Tohoku University, 6-6-11 Aramaki-aza Aoba, Aoba-ku, Sendai 980-8579, Japan Email: ye.songbo.t2@dc.tohoku.ac.jp

**Qingyuan Han -** School of Chemical and Biomolecular Engineering, The University of Sydney, Darlington, New South Wales, 2006, Australia. Email: qhan0135@uni.sydney.edu.au

**Jingwen Chi -** School of Chemical and Biomolecular Engineering, The University of Sydney, Darlington, New South Wales, 2006, Australia. Email: jchi2880@uni.sydney.edu.au

**Yuan Huang** School of Chemical and Biomolecular Engineering, The University of Sydney, Darlington, New South Wales, 2006, Australia. Email: yhua0254@uni.sydney.edu.au

**Heng Liu -** Advanced Institute for Materials Research (WPI-AIMR), Tohoku University, Sendai 980-8577, Japan. Email: heng.liu.e1@tohoku.ac.jp

**Di Zhang -** Advanced Institute for Materials Research (WPI-AIMR), Tohoku University, Sendai 980-8577, Japan. Email: di.zhang.a8@tohoku.ac.jp

**Xue Jia -** Advanced Institute for Materials Research (WPI-AIMR), Tohoku University, Sendai 980-8577, Japan. Email:

**Akichika Kumatani -** Department of Electrical and Electronic Engineering, Chiba Institute of Technology, Chiba 275-0016, Japan. Email: kumatani.akichika@p.chibakoudai.jp

## References

1. Papa, A. J., Propanal. In *Ullmann's Encyclopedia of Industrial Chemistry*, 2011.
2. Arntz, D.; Fischer, A.; Höpp, M.; Jacobi, S.; Sauer, J.; Ohara, T.; Sato, T.; Shimizu, N.; Schwind, H., Acrolein and Methacrolein. In *Ullmann's Encyclopedia of Industrial Chemistry*, 2007.
3. Kapil, N.; Coppens, M. O., Advances in the Hydroperoxidation of Propylene to Propylene Oxide (HOPO): from Nanoscale to Mesoscale and Macroscale. *Chemistry* **2025,** *31* (57), e01205.
4. Propylene Oxide (PO) Market Analysis: Industry Market Size, Plant Capacity, Production, Operating Efficiency, Demand & Supply, End-User Industries, Sales Channel, Regional Demand, Foreign Trade, Company Share, 2015–2036. https://www.chemanalyst.com/industry-report/propylene-oxide-po-market-755.
5. Nijhuis, T. A.; Makkee, M.; Moulijn, J. A.; Weckhuysen, B. M., The Production of Propene Oxide: Catalytic Processes and Recent Developments. *Ind. Eng. Chem. Res.* **2006,** *45* (10), 3447-3459.
6. Khatib, S. J.; Oyama, S. T., Direct Oxidation of Propylene to Propylene Oxide with Molecular Oxygen: A Review. *Catal. Rev.* **2015,** *57* (3), 306-344.
7. Teržan, J.; Huš, M.; Likozar, B.; Djinović, P., Propylene Epoxidation using Molecular Oxygen over Copper- and Silver-Based Catalysts: A Review. *ACS Catal.* **2020,** *10* (22), 13415-13436.
8. Anvari, S.; Medina, A.; Merchán, R. P.; Hernández, A. C., Sustainable solar/biomass/energy storage hybridization for enhanced renewable energy integration in multi-generation systems: A comprehensive review. *Renew. Sustain. Energy Rev.* **2025,** *223*.
9. Leow, W. R.; Lum, Y.; Ozden, A.; Wang, Y.; Nam, D. H.; Chen, B.; Wicks, J.; Zhuang, T. T.; Li, F.; Sinton, D.; Sargent, E. H., Chloride-mediated selective electrosynthesis of ethylene and propylene oxides at high current density. *Science* **2020,** *368* (6496), 1228-1233.
10. Dickens, C. F.; Kirk, C.; Nørskov, J. K., Insights into the Electrochemical Oxygen Evolution Reaction with ab Initio Calculations and Microkinetic Modeling: Beyond the Limiting Potential Volcano. *J. Phys. Chem. C* **2019,** *123* (31), 18960-18977.
11. Li, H.; Abraham, C. S.; Anand, M.; Cao, A.; Norskov, J. K., Opportunities and Challenges in Electrolytic Propylene Epoxidation. *J. Phys. Chem. Lett.* **2022,** *13* (9), 2057-2063.
12. Li, H.; Cao, A.; Nørskov, J. K., Understanding Trends in Ethylene Epoxidation on Group IB Metals. *ACS Catal.* **2021,** *11* (19), 12052-12057.
13. Sun, Y.; Hu, H.; Li, Y., Electro-Oxidation of Alkenes: A Green Approach Towards

Functionalized Oxygenates. *ChemCatChem* **2024,** *16* (13).

14. Chung, M.; Maalouf, J. H.; Adams, J. S.; Jiang, C.; Roman-Leshkov, Y.; Manthiram, K., Direct propylene epoxidation via water activation over Pd-Pt electrocatalysts. *Science* **2024,** *383* (6678), 49-55.

15. Pi, D.; Yang, X.; Chen, M.; Zhao, J.; Wang, J.; Wang, S.; Shih, W. C.; Xu, W.; Huang, Y.; Liu, B.; Li, X., In-situ spectroscopic insights into the dual-site synergistic electrocatalytic mechanism of propylene epoxidation over single-Ag-atom catalyst. *Nat. Commun.* **2025,** *17* (1), 691.

16. Lin, Y.; Li, H.; Miao, X.; Sun, Y.; Ren, H.; Yu, X.; Cui, W.; Wu, M.; Li, Z., V activated electro-epoxidation catalyst in membrane electrode assembly system for the production of propylene oxide. *Nat. Commun.* **2025,** *16* (1), 3113.

17. Koner, K.; Adams, J. S.; Miu, E. V.; Delgado-Kukuczka, S. P.; Jiang, C.; Kim, J. T.; Park, H.; Bui, J. C.; Oyala, P. H.; Li, Y.; Manthiram, K., Direct electrochemical propylene epoxidation over amorphized perovskite oxide in non-halogenated aqueous electrolyte. *Nat. Catal.* **2026**.

18. Li, D.; Sun, P.; Zhang, D.; Li, H.; Xu, H.; Cao, D., Unraveling the Potential-Dependent Selectivity of Propylene Electrooxidation: The Role of Electrochemistry-Induced Reconstruction. *J. Am. Chem. Soc.* **2025,** *147* (28), 24900-24912.

19. Liang, C.; Rao, R. R.; Svane, K. L.; Hadden, J. H. L.; Moss, B.; Scott, S. B.; Sachs, M.; Murawski, J.; Frandsen, A. M.; Riley, D. J.; Ryan, M. P.; Rossmeisl, J.; Durrant, J. R.; Stephens, I. E. L., Unravelling the effects of active site density and energetics on the water oxidation activity of iridium oxides. *Nat. Catal.* **2024,** *7* (7), 763-775.

20. Li, X.; Liu, L.; Ren, X.; Gao, J.; Huang, Y.; Liu, B., Microenvironment modulation of single-atom catalysts and their roles in electrochemical energy conversion. *Sci. Adv.* **2020,** *6* (39).

21. Ye, S.; Liu, F.; She, F.; Chen, J.; Zhang, D.; Kumatani, A.; Shiku, H.; Wei, L.; Li, H., Hydrogen Binding Energy Is Insufficient for Describing Hydrogen Evolution on Single-Atom Catalysts. *Angew. Chem. Int. Ed.* **2025,** *64* (23), e202425402.

22. Liu, F.; Zhang, D.; She, F.; Yu, Z.; Lai, L.; Li, H.; Wei, L.; Chen, Y., Mapping Degradation of Iron–Nitrogen–Carbon Heterogeneous Molecular Catalysts with Electron-Donating/Withdrawing Substituents. *ACS Catal.* **2024,** *14* (12), 9176-9187.

23. Xiong, L.; Zhan, S.; Ciano, L. D.; Li, S.; Liu, R.; Lu, S.; Yao, P.; Fu, X.; Yue, Q.,

Elucidating the Donor/Acceptor Regulatory Mechanism for $CO_2$ Electroreduction. *Adv. Mater.* **2026,** *38* (6), e12478.

24. Motagamwala, A. H.; Dumesic, J. A., Microkinetic Modeling: A Tool for Rational Catalyst Design. *Chem. Rev.* **2021,** *121* (2), 1049-1076.

25. Zhang, D.; Li, H., The potential of zero charge and solvation effects on single-atom M–N–C catalysts for oxygen electrocatalysis. *J. Mater. Chem. A* **2024,** *12* (23), 13742-13750.

26. Zhang, D.; Wang, Z.; Liu, F.; Yi, P.; Peng, L.; Chen, Y.; Wei, L.; Li, H., Unraveling the pH-Dependent Oxygen Reduction Performance on Single-Atom Catalysts: From Single- to Dual-Sabatier Optima. *J. Am. Chem. Soc.* **2024,** *146* (5), 3210-3219.

27. Wang, Y.; Zhang, D.; Sun, B.; Jia, X.; Zhang, L.; Cheng, H.; Fan, J.; Li, H., Divergent Activity Shifts of Tin-Based Catalysts for Electrochemical $CO_2$ Reduction: pH-Dependent Behavior of Single-Atom Versus Polyatomic Structures. *Angew. Chem. Int. Ed.* **2025,** *64* (8), e202418228.

28. Ye, S.; Wang, Y.; Liu, H.; Zhang, D.; Jia, X.; Zhang, L.; Zhang, Y.; Kumatani, A.; Shiku, H.; Li, H., Decoding pH-dependent electrocatalysis through electric field models and microkinetic volcanoes. *J. Mater. Chem. A* **2025,** *13* (44), 37821-37832.

29. Mathew, K.; Kolluru, V. S. C.; Mula, S.; Steinmann, S. N.; Hennig, R. G., Implicit self-consistent electrolyte model in plane-wave density-functional theory. *J. Chem. Phys.* **2019,** *151* (23), 234101.

30. Zhang, D.; Bao, Z.; Chu, Y.; Guo, Z.; Jia, X.; Jiang, Q.; Liu, H.; Liu, T.; Lu, T.; Lu, Y.; Shah, D. D.; Wang, Y.; Wang, Y.; Wang, Y.; Ye, S.; Ying, S.; Yu, Z.; Zhang, L.; Zhao, S.; Li, H., Digital catalysis platform as a gateway to big data and AI-powered innovations in catalysis. *Chem Catal.* **2026,** *6* (7).

31. Nørskov, J. K.; Bligaard, T.; Logadottir, A.; Kitchin, J. R.; Chen, J. G.; Pandelov, S.; Stimming, U., Trends in the Exchange Current for Hydrogen Evolution. *J. Electrochem. Soc.* **2005,** *152* (3).

32. Levey, K. J.; Frohlich, N. L.; Hardt, S.; de Kam, L. B. T.; Koper, M. T. M., The Origin of the Constant Phase Element Behavior of Pt(111) Near the Potential of Zero Charge. *ACS Electrochem.* **2026,** *2* (3), 693-704.

33. McCrum, I. T.; Koper, M. T. M., The role of adsorbed hydroxide in hydrogen evolution reaction kinetics on modified platinum. *Nat. Energy* **2020,** *5* (11), 891-899.

34. Kelly, S. R.; Heenen, H. H.; Govindarajan, N.; Chan, K.; Nørskov, J. K., OH Binding Energy as a Universal Descriptor of the Potential of Zero Charge on Transition Metal Surfaces. *J. Phys. Chem. C* **2022,** *126* (12), 5521-5528.
35. Cui, Y.; Wu, Y.; Ren, C.; Li, Q.; Ling, C.; Wang, J., Beyond Local Coordination: How Global Structure Engineers the Selectivity of Single Atom Catalysts for $CO_2$ Reduction. *Angew. Chem. Int. Ed.* **2026,** *65* (1), e19826.
36. Li, H.; Kelly, S.; Guevarra, D.; Wang, Z.; Wang, Y.; Haber, J. A.; Anand, M.; Gunasooriya, G. T. K. K.; Abraham, C. S.; Vijay, S.; Gregoire, J. M.; Nørskov, J. K., Analysis of the limitations in the oxygen reduction activity of transition metal oxide surfaces. *Nat. Catal.* **2021,** *4* (6), 463-468.